\documentstyle[aps,prl,amsmath,draft]{revtex}
\begin{document}
\draft

\title{Continuously-variable survival exponent for random walks 
with movable partial reflectors}

\author {Ronald Dickman$^{1,*}$ and Daniel ben-Avraham$^{2,\dag}$}
\address{
$^1$Departamento de F\'\i sica, ICEx,
Universidade Federal de Minas Gerais, Caixa Postal 702,
30161-970, Belo Horizonte - MG, Brasil\\
$^2$Physics Department and Center for Statistical Physics (CISP),
Clarkson University, Potsdam, New York 13699-5820}
\date{\today}

\maketitle
\begin{abstract}
We study a one-dimensional lattice random walk
with an absorbing boundary at the origin and a
movable partial reflector. 
On encountering the reflector, at site $x$, the walker is reflected 
(with probability $r$) to $x-1$ and the reflector is simultaneously 
pushed to $x+1$.  
Iteration of the transition matrix, and asymptotic analysis of
the probability generating function show that
the critical exponent $\delta$ governing the survival probability 
varies continuously between 1/2 and 1 as $r$ varies between 0 and 1.  
Our study suggests a mechanism for nonuniversal kinetic critical behavior,
observed in models with an infinite number of absorbing configurations.



\end{abstract}

Random walks find application in virtually every area of physics
\cite{barber,vankampen,sokal,weiss,dba-havlin}.  Random walks in the
presence of traps or absorbing boundaries, and/or reflectors, are much
studied as models of exciton recombination \cite{kopelman},
diffusion-limited reactions \cite{dba-havlin}, and polymer-surface
interactions \cite{polysurf}.

In this Letter we study an unbiased random walk with an absorbing 
boundary at 
the origin, and a movable, partial reflector.  At each encounter between 
walker and reflector, the latter moves one step to the right
(i.e., to a site not yet visited by the walker), while the walker is 
reflected to its previous position with probability $r$.  Thus
the reflector hampers the advance of the walker into new
territory, but does not otherwise influence its motion.
(Note that in the limit $r=0$ the reflector
has no effect, but it does mark the {\em span} of the walk.) We are
primarily interested in the effect of the reflector on the asymptotic
scaling properties of the walk.

The process defined above admits various physical interpretations.
One is in terms of adlayer growth, with 
deposition and evaporation at the edge of the adlayer (but not in the
bulk).  The substrate adsorption sites are originally in a 
``non-activated" state, with a low sticking probability for
incident molecules, but after a first adsorption attempt (at a
step edge) the substrate site becomes activated, with a high
sticking probability.  Biological interpretations are also possible, 
e.g., of the advance of a bacterial colony in a growth medium, with a 
preliminary contact facilitating expansion into new regions, or,
similarly, the spread of a political viewpoint in an initially
skeptical population.
Our results are of interest as an example of
nonuniversality (a continuously-variable critical exponent) in a model that
allows an exact asymptotic analysis.

A further motivation for our study is provided by absorbing-state phase 
transitions, an area of great current interest in nonequilibrium 
statistical physics \cite{marro,hinrichsen},
in which a many-particle system, such as directed percolation, becomes
trapped in a configuration allowing no further evolution.  Continuous
transitions to an absorbing state have been invoked in models of
epidemics \cite{harris}, catalysis \cite{zgb}, and the transition
to turbulence \cite{pomeau,chate}, to cite but a few examples.
The connection between absorbing-state phase transitions and random
walks lies in {\it compact} directed percolation (CDP) \cite{cdp}.
CDP is defined on a discrete space-time $(x,t)$
with time-slices corresponding to diagonals of a square lattice, so that
the neighbors of site $(x,t)$ at the previous time are $(x\!-\!1,t\!-\!1)$ 
and $(x\!+\!1,t\!-\!1)$.  Each site is either occupied or vacant.  
If $(x,t-1)$ has $n$ occupied neighbors,
then $(x,t)$ is vacant (occupied) with probability 1 if $n=0$ (2),
and is occupied with probability $p$ if $n=1$.  
CDP exhibits an absorbing-state phase 
transition at $p=1/2$.  Consider an initial
state of a single occupied site in an otherwise empty lattice.  
The boundaries between the occupied region (descended from the initial
seed particle) and the outer vacant regions follow simple random
walks, which are unbiased if $p=1/2$.  Thus the {\it length} $X(t)$ of the
occupied region, being the distance between two random walks, 
is also a random walk, with an absorbing boundary at
the origin.  For $p<1/2$ ($>1/2$) $X(t)$ is attracted to (repelled by)
the origin.
For $p=1/2$, $X(t)$ is unbiased, and well known results for
random walks \cite{barber,weiss,feller} imply that the survival
probability decays for long times $ \sim t^{-\delta}$, with $\delta=1/2$.

The motivation for introducing mobile reflectors in CDP (and thus
in the simple random walk studied here) arises from the
puzzling behavior of models that can become trapped in one of an
{\em infinite} number of absorbing configurations (INAC)
\cite{pcp,mendes,inas}. Anomalies in critical spreading for INAC, such as
continuously variable critical exponents, have been
traced to a long memory in the dynamics of the order parameter, $\rho$,
due to coupling to an auxiliary field that remains frozen in regions where
$\rho = 0$ \cite{inas,mgd}.  INAC appears to be particularly relevant
to the transition to spatio-temporal chaos, as shown in a recent study
of a coupled-map lattice with `laminar' and `turbulent' states, which 
revealed continuously-variable spreading exponents \cite{bohr}.

Grassberger, Chat\'e and Rousseau (GCR) \cite{gcr}
proposed that spreading in systems with INAC could be understood
by studying a model with a {\it unique} absorbing configuration,
but in which the spread of activity to previously inactive regions 
is hampered (or facilitated).
Our model represents a two-fold
simplification of the GCR model: first, the appearance of inactive sites
within a string of active ones is prohibited (analogous to going from DP
to the more restrictive CDP process); second, we study a single
random walker rather than the pair needed to describe CDP \cite{note1}.  
We believe, nonetheless, that our model captures the essential physics 
underlying anomalous critical behavior in models with INAC.

We study an unbiased, discrete-time random walk on the nonnegative
integers, $x_t = 0, 1, 2,...$, with $x\!=\!0$ absorbing.  
Initially the walker is at $x_0\!=\!1$ and the reflector, whose position we 
denote by $R_t$, is at $R_0\!=\!2$.  Each time the walker steps to the site
occupied by the reflector, it is reflected back to $R-1$ with
probability $r$ (and remains at $R$ with probability $1-r$),
while the reflector moves (with probability 1) to $R+1$.
Evidently the process $x_t$ is non-Markovian, since the transition
probability into a given site depends on whether it has been visited
before.  We can transform the model to a Markov process by
enlarging the state space to include the reflector position; it is
convenient to introduce the variable $y_t=R_t \!-\!1$ for this purpose.
The process $(x,y)$ is restricted to the wedge between $x=0$ (absorbing) 
and $x=y$, with transitions from $y$ to $y+1$ allowed only from the
diagonal $x=y$.  At this point it is useful to include a further
generalization of our model, by assigning a probability $p'$ for
the walker to jump to the right (and $q'=1\!-\!p'$ to jump to the left) 
when on the diagonal; these transition probabilities are summarized
in Fig.~1.   For $p' < 1/2$ the walker experiences an additional
impediment to visiting new territory, while for $p'(1-r) > 1/2$
the `reflector' effectively becomes an accelerator, drawing the walker
forward to a previously unvisited site.

The nonzero transition probabilities for the Markov chain are:

\begin{eqnarray}
W[(x,y) \;\; \to \;\; (x\!\pm\!1,y)] &=&  1/2, \;\;\;\; x=1,...,y\!-\!1
\\ \nonumber
W[(y,y) \;\; \to \;\; (y,y+1)] &=&  p'r
\\ \nonumber
W[(y,y) \to (y\!+\!1,y\!+\!1)] &=&  p'(1\!-\!r)
\\ \nonumber
W[(y,y) \;\; \to \;\; (y\!-\!1,y)] &=&  1-p' \;.
\label{trprb}
\end{eqnarray}
The probability $P(x,y,t)$ follows the master equation
\begin{equation}
P(x,y,t+1)=\case{1}{2}P(x\!-\!1,y,t)+\case{1}{2}P(x\!+\!1,y,t),
  \;\;\;\ x=1,2,\dots,y\!-\!2;\quad y\geq3\;,
\label{recrel}
\end{equation}
with $P(0,y,t)=0$, representing the absorbing boundary at $x=0$. Letting
$D(y,t) \equiv P(y,y,t)$, the boundary conditions along the diagonal are:

\begin{eqnarray}
&& P(y\!-\!1,y,t\!+\!1)=\case{1}{2}P(y\!-\!2,y,t)+p'rD(y\!-\!1,t)
 +(1\!-\!p')D(y,t)\;,
\\ \nonumber
&& D(y,t\!+\!1)=\case{1}{2}P(y\!-\!1,y,t)+p'(1\!-\!r)D(y\!-\!1,t)\;,
\end{eqnarray}
for $y\geq3$.  For $y=1$, 2 the evolution equations depend on the
initial condition; here $P(x,y,0) =
\delta_{x,1}\delta_{y,1}$~\cite{note2}.

Define the generating function
$\hat{P}(x,y,z)=\sum_{t=0}^{\infty}z^t\,P(x,y,t)$
(and similarly for $D$, etc.).  $\hat{P}$ satisfies
\begin{equation}
\label{evolution}
z^{-1}\hat{P}(x,y)=\case{1}{2}\hat{P}(x\!-\!1,y)
 +\case{1}{2}\hat{P}(x\!+\!1,y),
  \qquad x=1,2,\dots,y-2;\quad y\geq3\;.
\end{equation}
(we drop the argument $z$ for brevity),
subject to the boundary conditions

\begin{eqnarray}
\label{bc.x=0}
&& \hat{P}(0,y)=0\;, \\ 
\label{bc.x=y-1}
&&z^{-1}\hat{P}(y\!-\!1,y)=\case{1}{2}\hat{P}(y\!-\!2,y)
 +p'r\hat{D}(y\!-\!1)+
(1\!-\!p')\hat{D}(y)\;,\\
\label{bc.x=y}
&&
z^{-1}\hat{D}(y)=\case{1}{2}\hat{P}(y\!-\!1,y)
 +p'(1\!-\!r)\hat{D}(y\!-\!1)\;. 
\end{eqnarray}
The solution of Eqs.~(\ref{evolution}) and (\ref{bc.x=0}) is
\begin{equation}
\label{P}
\hat{P}(x,y)=\hat{C}(y)(\lambda_+^x-\lambda_-^x)\;;\qquad
  \lambda_{\pm}= z^{-1} \pm\sqrt{z^{-2}-1}\;,
\end{equation}
with $\hat{C}(y)$ yet to be determined. Noting that
$\lambda_+=\lambda_-^{-1} \equiv \lambda$, we
define $\lambda(x) \equiv \lambda^x-\lambda^{-x}$; thus
$\hat{P}(x,y)=\hat{C}(y)\lambda(x)$.  
Note also the recurrence relation for integer $x$:
$\lambda(x\!-\!1)-2z^{-1}\lambda(x)+\lambda(x\!+\!1)=0$.

>From Eq.~(\ref{bc.x=y}) we find
\begin{equation}
\label{D}
\hat{D}(y)=\omega^{y-2} \hat{D}(2)
  +\case{1}{2} z \omega^y \sum_{y'=3}^y\omega^{-y'}\hat{P}(y'\!-\!1,y')  \;,\qquad
  \omega=p'(1\!-\!r)z\;.
\end{equation}
It follows that $\hat{D}(y)-\omega\hat{D}(y\!-\!1)
=\case{z}{2}\hat{C}(y)\lambda(y\!-\!1)$. 
Therefore, subtracting from Eq.~(\ref{bc.x=y-1}) $\omega$ times 
the corresponding equation for $y\!-\!1$, and using the recursion relation
for $\lambda(x)$, we find
\begin{equation}
\frac{\hat{C}(y)}{\hat{C}(y\!-\!1)} 
 = p'\frac{r\lambda(y\!-\!2)+(1\!-\!r)\lambda(y\!-\!1)}
  {(p'\!-\!\case{1}{2})\lambda(y\!-\!1)+\case{1}{2}\lambda(y\!+\!1)}\;,
\end{equation}
which, noting that $\hat{C}(2) = \hat{P}(1,2)/\lambda(1)$, provides
the complete solution for the probability generating function.

To determine the  survival probability
$S(t)=\sum_{y=0}^{\infty}\sum_{x=0}^yP(x,y,t)$ as $t \to \infty$,
we analyze the singular behavior of $\hat{S}(z)$ as $z \to 1$.
In this limit $\lambda = 1 + \sqrt{2\epsilon} + {\cal O}(\epsilon) $, 
where $\epsilon = 1-z$, and we find the dominant term to be \cite{later}
\begin{equation}
\hat{S} \sim  
\case{1}{\sqrt{2(1-z)}}\sum_{y=3}^{\infty}\hat{C}(y)[\lambda(y/2)]^2 \;,
\label{sasy}
\end{equation}
which converges for $z < 1$.  Note that 
$\lambda(n) \sim 2n\sqrt{2\epsilon}$
for $n\epsilon \ll 1$, while in the opposite limit 
$\lambda(n) \sim \lambda^n$.
For $n\epsilon \ll 1$ we may thus use the $z \to 1$ limiting expression
for $\hat{C}(n)$, that is,
\begin{equation}
\hat{C}_0(y)=\frac{\hat{P}(1,2)}{2\sqrt{2(1-z)}}
  \frac{\Gamma(2+1/p')}{\Gamma(2-r)}\frac{\Gamma(y-r)}{\Gamma(y+1/p')}\;.
\end{equation}
We then approximate the summand in Eq.~(\ref{sasy}) by
\begin{equation}
f_n = 
\begin{cases}
 2 n^2 \hat{C}_0(n) \epsilon, & n\epsilon \leq 1\;, \\
 2 N^2 \hat{C}_0(N) \epsilon \,a^{n-N}\lambda^{n-N}, & n\epsilon > 1 \;,
	\end{cases} 
\end{equation}
where $N = 1/\sqrt{\epsilon} $ rounded to an integer, and
\begin{equation}
a=p'\frac{r+(1-r)\lambda}{(p'-1/2)\lambda+(1/2)\lambda^3}\;.
\end{equation}

Since $\hat{S}$ and $F(z) \equiv \sum_{n=0}^\infty f_n $ have the same 
radius of convergence and $\lim_{n \to \infty} \hat{C}(n)
[\lambda(n/2)]^2/f_n $ is finite, an Abelian theorem implies that the
singular behavior of $\hat{S}$ is the same as that of
$F(z)/\sqrt\epsilon$ \cite{weiss}.  Write
$F(z)/\sqrt\epsilon = 2\epsilon^{1/2} \sum_{n=0}^N n^2 \hat{C}_0(n)
+2\epsilon^{1/2}N^2 \hat{C}_0(N)\sum_{n=N+1}^\infty (\lambda
a)^{n-N}$. We approximate the first sum by an integral; using 
$\hat{C}_0 (n) \sim \epsilon^{-1/2}n^{-r-1/p'}$ we then find that the 
first term
$\sim \epsilon^{(r+1/p' - 3)/2}$.  The same holds for the second term
when we note that $1-a\lambda\sim\epsilon^{1/2}$ and
$\hat{C}_0(N)\sim\epsilon^{-\frac{1}{2}(1+r+1/p')}$.  Thus $\hat{S} \sim
(1-z)^{\delta - 1} $ (plus less singular terms) as $z \to 1$, where
$\delta = (r + 1/p' - 1)/2$.  Finally, noting that for large $n$, the
coefficient of $z^n$ in $(1-z)^{\delta - 1}$ is 
$n^{-\delta}/\Gamma(1-\delta)$, we have that the survival probability
decays asymptotically $\sim t^{-\delta}$;  in particular, for $p'=1/2$
we have $S(t) \sim t^{-(1+r)/2}$.
Note also that the larger $p'$ is, the slower the decay of the survival
probability, as expected.

While the amplitude of the leading term in $S(t)$ can be evaluated
exactly for $p'=1/2$ and $r=0$ or 1~\cite{later}, we
have not found a closed-form expression in the general case.
Numerical evaluation of the
exact expression for $\hat{S}$ does, however, lead to an amplitude in
perfect agreement with that found in the transition matrix analysis.

We turn now to results obtained via numerical iteration of
the transition matrix defined in Eq.~(\ref{trprb}).
We studied the process to a maximum
time of up to $3 \times 10^5$, quite sufficient to observe asymptotic
behavior.  In Fig.~2 we show the survival probability for $p'=1/2$
and various values of $r$.  In each case there is a power-law decay,
with the decay exponent $\delta$ varying continuously with $r$.  
(The plateaus in $S(t)$ at short times reflect that for $r=0$
the walker can only reach the origin on odd-numbered steps.
Similarly, for $r=1$ the walker cannot reach the origin on the third step.) 
We verify that the mean position $\langle x_t \rangle \sim t^{1/2}$
(and that $\langle x_t^2 \rangle \sim t$; the averages are over 
surviving walks) independent of $r$.  The amplitudes for the displacement
and its second moment do decrease with increasing $r$.

To determine the decay exponent precisely, we study the local slope 
$\delta(t)$, given by a least-squares linear fit
to the $\ln S$ data for a set of 11 equally-spaced values (increments of
0.05) of $\ln t$.  Plots of $\delta (t)$ vs.~$t^{-1}$ lead to
asymptotic ($t \to \infty$) values confirming $\delta = (1+r)/2$ to
better than one part in 2000.  It is possible to 
improve the analysis further by determining the leading correction to 
scaling in an expansion of the form:
$ S(t) \simeq A t^{-\delta} [ 1 - B t^{-\Delta_1} + \cdots]$.
For $r=0$ we find $A=0.7979$, $B=0.78$ and $\Delta_1 = 1.00$,
in accord with the exact expansion 
$P(t) \simeq \sqrt{2/(\pi t)}[ 1- (3/4)t^{-1} + \cdots]$.
For $r=1$ our data yield $A=1.0000$ (in agreement with the exact result),
$B=1.332$ and $\Delta_1 = 1.00$.
In these two cases, plotting $\delta (t) $ versus $t^{-\Delta_1} = t^{-1}$ 
we obtain $\delta = 0.5000.$ and 0.999985, respectively.  For $r$
in the interval [0.1,0.8] we find a much smaller correction to scaling
exponent, $\Delta_1 \simeq 0.54$, while for 
$r=0.9$, $\Delta_1 \simeq 0.62$.
Plotting $\delta(t) $ versus $t^{-\Delta_1}$ (see Fig.~3), 
we obtain exponents that agree with the theoretical
value to better than one part in $2 \times 10^4$.
The attractive simplification, $\Delta_1 = 1/2$
for $r\neq0$, 1, yields essentially the same values for $\delta$, 
and is in fact to be expected,
given that generic corrections to $\hat{S}$ should be 
$\propto (1-z)^{\delta - 1 + \frac{1}{2}m}$  $(m = 1, 2,\dots)$.  

The case of compact directed percolation is somewhat more complicated
than the random walk problem analyzed above, as we must now keep
track of three variables, viz., the distance $X$ between the two walkers,
and the respective distances between the walkers and the associated
reflectors \cite{note1}.  Simulation results for CDP with movable
partial reflectors again show $\delta$ varying continuously with $r$,
but over a somewhat wider range: for $r=1$, $\delta = 1.158(3)$.

In one-dimensional spreading models (contact process \cite{harris}, pair contact
process \cite{pcp}, etc.)  the survival and spread of activity
may again be described in terms of the position of 
two points that bound the
active region, as in CDP.  But here the boundaries 
of the active region do not
follow simple random walks: their dynamics involves a variable step size 
(due to gaps in the distribution of active sites), and
a significant memory (the step-size distribution depends on the history of
the previous step directions).  
It seems reasonable, nevertheless, to expect
that varying the probability for advancing into virgin territory will 
change the scaling of the survival probability, 
just as observed here.  A detailed
investigation of this issue is a high priority for future work.

In summary, we have uncovered the rather remarkable property of a 
continuously varying
critical exponent governing the survival probability of a random walk
with an absorbing boundary,
in the presence of a movable partial reflector.
Our results suggest a new approach to understanding
nonuniversal spreading in models with an infinite number of absorbing 
configurations.

\acknowledgements

We thank L.~S.~Schulman for stimulating discussions.
R.D. is supported by CNPq. D.b.-A. thanks the NSF (PHY-9820569) for
support.

\noindent 
$^*${\small electronic address: dickman@cedro.fisica.ufmg.br } \\
$^{\dag}${\small electronic address: benavraham@clarkson.edu } \\

\noindent Figure Captions
\vspace{1em}

FIG. 1. Transition probabilities in the (x,y) plane.
\vspace{1em}

FIG. 2. Survival probability versus time for random walk with a movable
partial reflector, $p'=1/2$.  Reflection probability (top to bottom) $r = 0,
0.1,\dots,1$.
\vspace{1em}

FIG. 3. Local slope $\delta (t)$ for 
$p'=r=1/2$.  The inset shows the same data plotted versus $1/t$.


\begin{thebibliography}{99}
\bibitem{note1}
     Detailed results for compact directed percolation will be reported
     elsewhere.
\end{thebibliography}
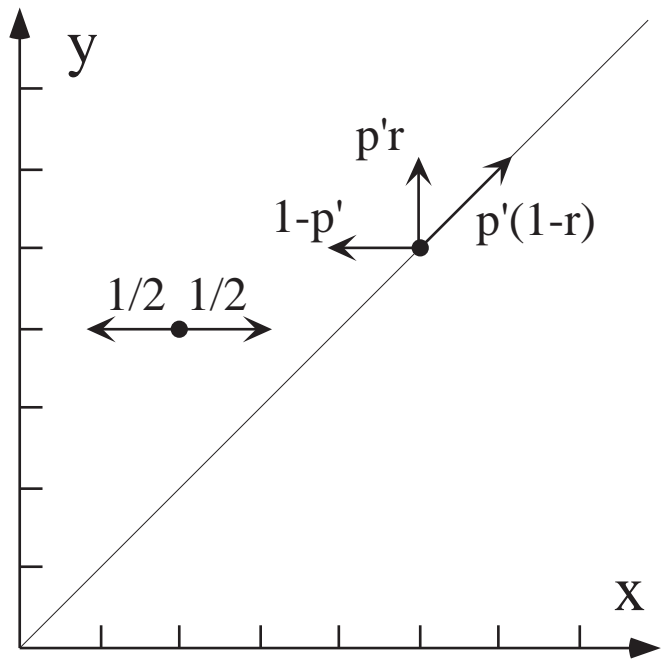
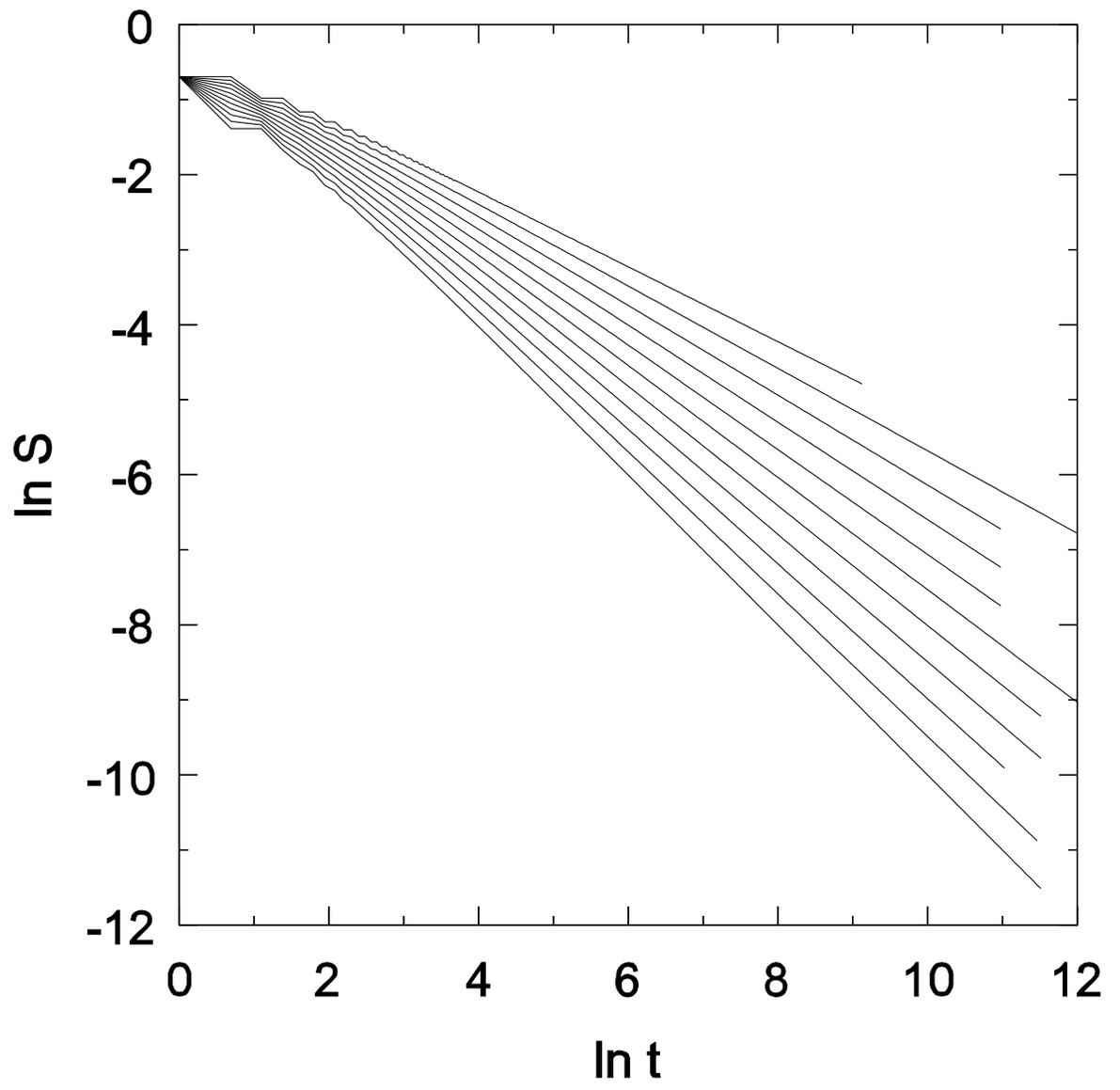
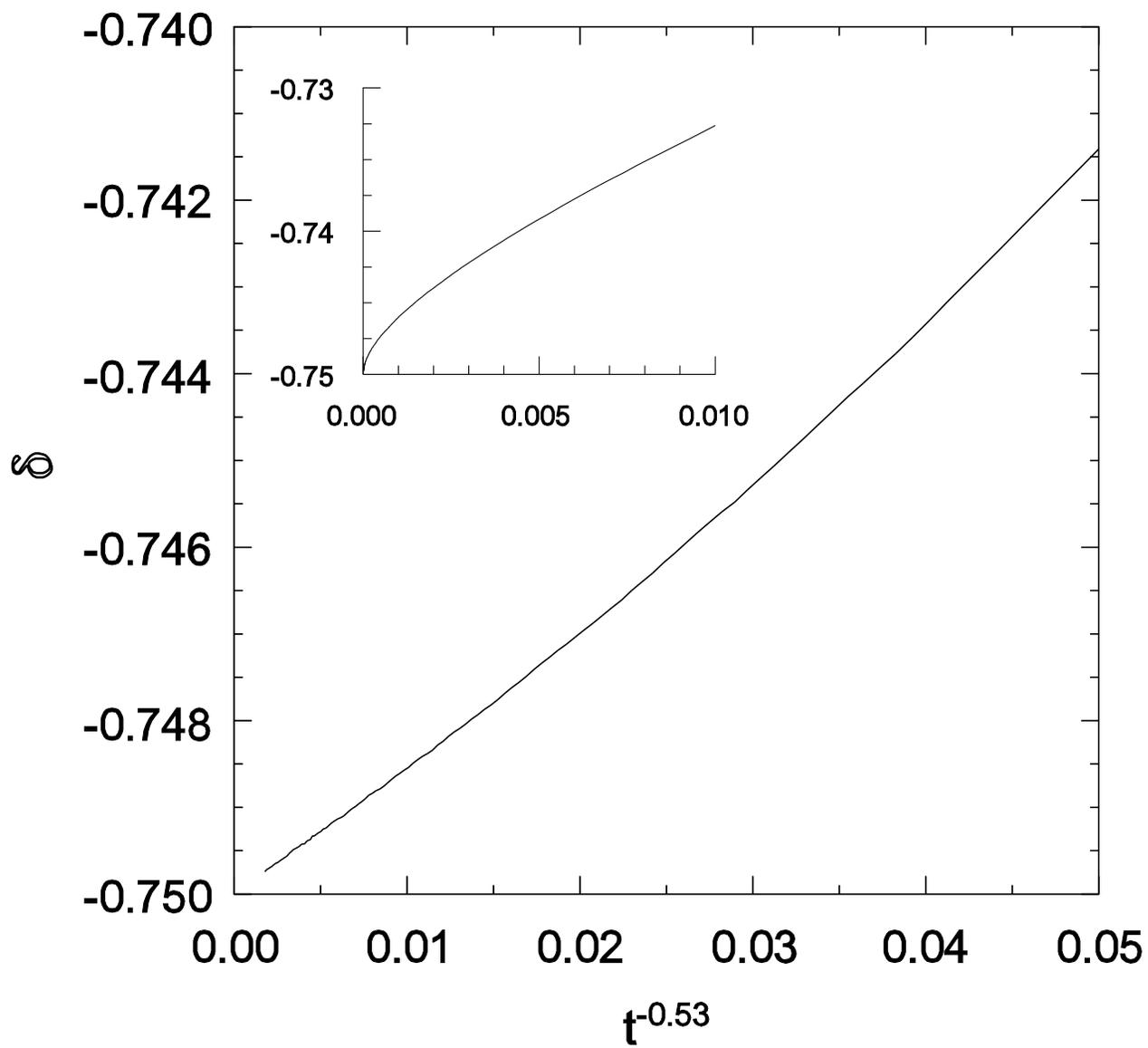
\end{document}